\documentclass[fleqn,usenatbib]{mnras}

\usepackage{textcomp}
\usepackage{ae,aecompl}
\usepackage{datatool}

\usepackage{graphicx}	
\usepackage{amsmath}	
\usepackage{amssymb}	
\newcommand\textlcsc[1]{\textsc{\MakeLowercase{#1}}}

\title[Streams and Milky Way mass]{The effects of dynamical substructure on Milky Way mass estimates from the high velocity tail of the local stellar halo}

\author[R. J. J. Grand et al.]{\parbox[t]{\textwidth}{
Robert J. J. Grand,$^{1}$\thanks{E-mail: grand@mpa-garching.mpg.de}, Alis J. Deason$^2$, Simon D. M. White$^1$, Christine M. Simpson$^{3,4}$, Facundo A. G\'{o}mez$^{5,6}$, Federico Marinacci$^7$, R\"{u}diger Pakmor$^1$}
\vspace{10pt}
\\ 
$^1$Max-Planck-Institut f\"{u}r Astrophysik, Karl-Schwarzschild-Str. 1, 85748 Garching, Germany\\
$^2$Institute for Computational Cosmology, Department of Physics, Durham University, South Road Durham DH1 3LE, UK \\
$^3$Department of Astronomy \& Astrophysics, The University of Chicago, Chicago, IL 60637, USA\\
$^4$Enrico Fermi Institute, The University of Chicago, Chicago, IL  60637, USA\\
$^5$Instituto de Investigaci{\'o}n Multidisciplinar en Ciencia yTecnolog{\'i}a, Universidad de La Serena, Ra{\'u}l Bitr{\'a}n 1305, La Serena, Chile\\
$^6$Departamento de F{\'i}sica y Astronom{\'i}a, Universidad de LaSerena, Av. Juan Cisternas 1200 N, La Serena, Chile\\
$^7$Department of Physics \& Astronomy, University of Bologna, via Gobetti 93/2, 40129 Bologna, Italy
}

\date{Accepted XXX. Received YYY; in original form ZZZ}

\pubyear{2019}

\begin{document}
\label{firstpage}
\pagerange{\pageref{firstpage}--\pageref{lastpage}}
\maketitle

\begin{abstract}
We investigate the impact of dynamical streams and substructure on estimates of the local escape speed and total mass of Milky Way-mass galaxies from modelling the high velocity tail of local halo stars. We use a suite of high-resolution, magneto-hydrodynamical cosmological zoom-in simulations, which resolve phase space substructure in local volumes around solar-like positions. We show that phase space structure varies significantly between positions in individual galaxies and across the suite. Substructure populates the high velocity tail unevenly and leads to discrepancies in the mass estimates. We show that a combination of streams, sample noise and truncation of the high velocity tail below the escape speed leads to a distribution of mass estimates with a median that falls below the true value by $\sim 20 \%$, and a spread of a factor of 2 across the suite. Correcting for these biases, we derive a revised value for the Milky Way mass presented in Deason et al. of $1.29 ^{+0.37}_{-0.47} \times 10^{12}$ $\rm M_{\odot}$.
\end{abstract}




\section{Introduction}

The mass of the Milky Way (MW) is one of the most fundamental, basic parameters of our Galaxy, yet remains uncertain by about a factor of 2. Historically, most mass estimates are inferred from stellar tracers of the Galactic potential, which include the kinematics of halo stars \citep[e.g.][]{LT90,PSB14,MFC18}, the kinematics of satellite galaxies \citep[e.g.][]{CCD19} and globular clusters \citep[e.g.][]{WMS18,PH19}, modelling of stellar streams \citep[e.g.][]{MIM18} and a timing argument applied to the orbit of the distant satellite Leo I \citep{ZOS89,LW08,PG16}. Weakly bound halo stars have long been thought to be valuable probes of the Galactic potential at large radii, and have, since the 1980s, been used to infer the local Galactic escape speed \citep[e.g.][]{CO81}. As more of these stars became observable, statistical techniques were developed to constrain the local escape speed from parametrized modelling of the velocity distribution \citep{LT90}. The most recent applications of this approach utilize data sets with precise 3-dimensional velocities and distances, and yield mass determinations of $\sim 1.00 ^{+0.31}_{-0.24} \times 10^{12}$ $\rm M_{\odot}$ \citep[][D19 hereafter]{DFB19}. These results compare well with estimates based on other tracers provided by the {\it Gaia} second data release, which range from $\sim 1-1.5 \times 10^{12}$ $\rm M_{\odot}$. 

The popular method of \citet{LT90} relies on the assumption that the stellar distribution function is smooth, well mixed and has a high velocity tail which extends to the local escape speed with the form of a power-law, $f \propto (v_e- v)^k$. However, the stellar halo contains a lot of highly structured features, including the well known Sagittarius stream, that wraps more than one full ring around the sky \citep[e.g.][]{SHD17,FML19}. In addition to this dominant stream, the local halo contains a wealth of smaller scale, clumpy substructure \citep{HVB17,KHV18}, some of which may arise from recently accreted progenitors \citep[e.g. $\omega \rm Cen$, see][]{MEB18b,IMM19}. This range of structure in the stellar halo is supported by recent cosmological simulations \citep[][]{HWS03,GHC13,cooper15a,MGG19,SSG19}, which show that progenitors of the stellar halo can produce streams in the local phase space distribution, particularly if they undergo several pericentric passages as dynamical friction drags them to the galaxy centre. It is reasonable to expect that such a distribution would: i) be populated unevenly; ii) vary as a function of position within the Galaxy; and iii) lead to potentially significant biases in the escape speed estimate and therefore the total mass estimate of the MW.

To understand the effects of substructure on MW mass estimates, we use a suite of high-resolution cosmological simulations of the formation of MW-mass galaxies in the $\Lambda$CDM paradigm \citep[][]{GGM17}. The high resolution allows for exquisite sampling of the gravitational potential such that a range of stellar substructure in local volumes is captured. We apply the analysis of D19 to local volumes centred on four equidistant solar positions in the disc of each simulated galaxy, and derive escape speeds and MW mass estimates. We find that the estimates for the escape speed and total mass vary significantly between positions and individual haloes, and in some cases deviate strongly from the true values. We link these trends with dynamical substructure and the high velocity tail, quantify the biases and discuss the implications for the MW.

\section{Simulations}

We analyse local samples of accreted halo stars in a suite of high resolution, cosmological magneto-hydrodynamical simulations for the formation of the MW \citep[from the \textlcsc{Auriga} project,][]{GGM17,GHF18}. Our suite comprises simulations from two different resolution levels: level 3 ($\sim 6 \times 10^3$ $\rm M_{\odot}$ per baryonic element; softening length of $184$ pc after $z=1$) and level 4 ($\sim 5 \times 10^4$ $\rm M_{\odot}$ per baryonic element; softening length of $369$ pc after $z=1$), to maximize statistics. The haloes range between $0.5$-$2\times 10^{12} \rm M_{\odot}$ in total mass ($M_{200}$), which we define as the mass contained inside the radius at which the mean enclosed mass volume density equals 200 times the critical density for closure. Each halo was initially selected from the $z=0$ snapshot of a parent dark matter only cosmological simulation of comoving periodic box size 100 Mpc, with the standard $\Lambda$CDM cosmology. The adopted cosmological parameters are $\Omega _m = 0.307$, $\Omega _b = 0.048$, $\Omega _{\Lambda} = 0.693$ and a Hubble constant of $H_0 = 100 h$ km s$^{-1}$ Mpc$^{-1}$, where $h = 0.6777$, taken from \citet{PC13}. At $z=127$, the resolution of the dark matter particles of this halo is increased and gas is added to create the initial conditions of the zoom, which is evolved to present day with the magneto-hydrodynamics code \textlcsc{AREPO} \citep{Sp10}. 

The simulations include a comprehensive galaxy formation model \citep[][]{GGM17}, which treats many relevant processes, including gravity, gas cooling, star formation, mass and metal return from stellar evolutionary processes, energetic stellar and AGN feedback, and magnetic fields. Large cosmological box simulations \citep[e.g.][]{PNH18,NPS18} have demonstrated that a similar galaxy formation model reproduces many of the global properties of the observed galaxy population, such as the galaxy stellar mass function, galaxy sizes, the cosmic star formation rate density and galaxy morphological mix. The combination of a globally successful model with the ability to resolve internal galactic structures, such as bars and spiral arms, makes the simulation suite ideal to study Galactic dynamics within the full cosmological setting.

\section{Estimating the escape speed and mass of the Galaxy}
\label{meth}

The mass estimation procedure we test is based on that of D19 and consists of two parts: i) the estimation of the local escape speed from local, accreted star particles, which is sensitive to the outer halo; and ii) the estimation of the total mass using the circular velocity at the Solar radius as an additional constraint. We briefly summarise the method below, and refer the reader to D19 for more details.

For a given solar position, we select all accreted star particles within a 3 kpc sphere. We assume that the velocity distribution follows a power law: $ f \propto (v-v_e)^k$, where $v$ is the total velocity of the particle in the Galactic rest frame, $v_e$ is the escape velocity at the position of the star and $k$ is the power law slope. We further assume that the escape velocity varies as a function of galactocentric radius as \citep[e.g.][]{WBC17}: $v_e = v_{e,0}(r/r_0)^{\gamma/2}$, where $r_0 = 8$ kpc is the solar radius, and $v_{e,0}$ is the escape speed at the position of the Sun. We use a log-likelihood analysis to evaluate probability density functions for $v_{e,0}$,  $\gamma$ and $k$, on uniform grids in the range $v_{e,0} \in [300,900]$, $\gamma \in [0,1]$ and $k \in [k_{\rm min},k_{\rm max}]$. Here, $k_{\rm min}$ and $k_{\rm max}$ are defined to be equal to $k_{\rm av} \pm 1$, where $k_{\rm av}$ is measured from all accreted star particles in the radial range $4 < r/\rm kpc < 12$ for each simulation, which mimics the construction of the prior on $k$ adopted by D19 for the MW. For further consistency, we adopt the prior $P(v_{e,0}) \propto 1/v_{e,0}$.

\begin{figure*}
\centering
\hspace{0.9cm}
\includegraphics[scale=0.8,trim={0.95cm 0 0.5cm 0.5cm},clip]{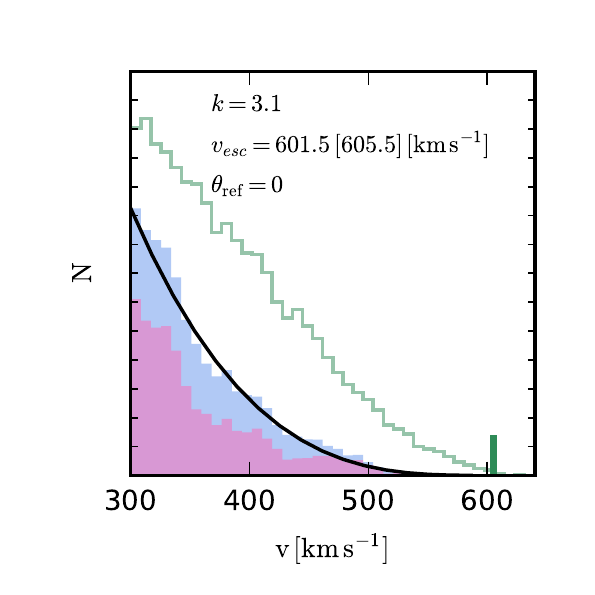}
\includegraphics[scale=0.8,trim={1.1cm 0 0.5cm 0.5cm},clip]{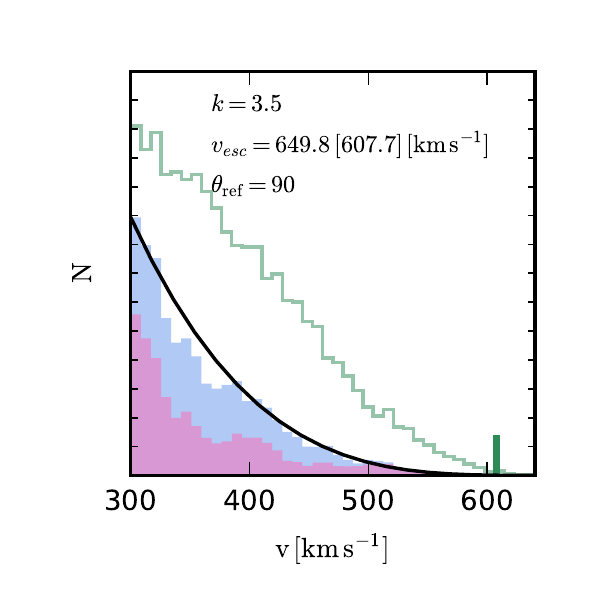}
\includegraphics[scale=0.8,trim={1.1cm 0 0.5cm 0.5cm},clip]{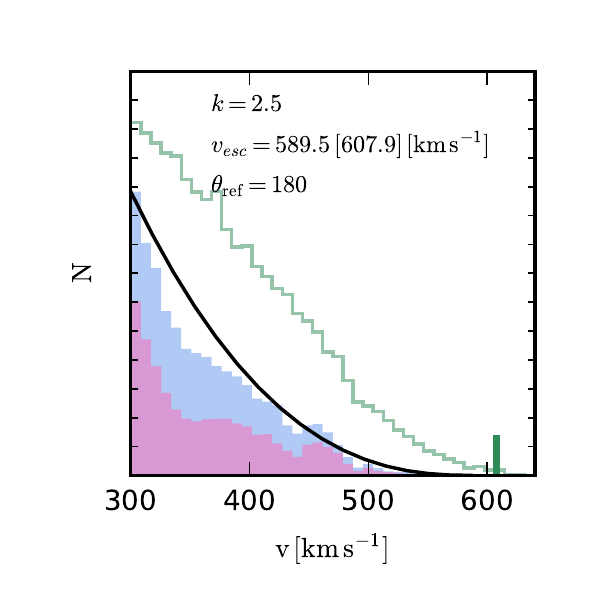}
\includegraphics[scale=0.8,trim={1.1cm 0 0.5cm 0.5cm},clip]{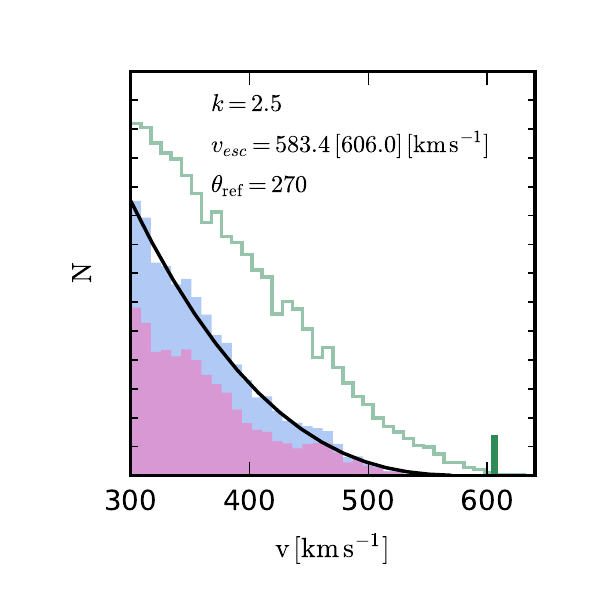}\\
\includegraphics[scale=0.8,trim={0.35cm 0 0 0.5cm},clip]{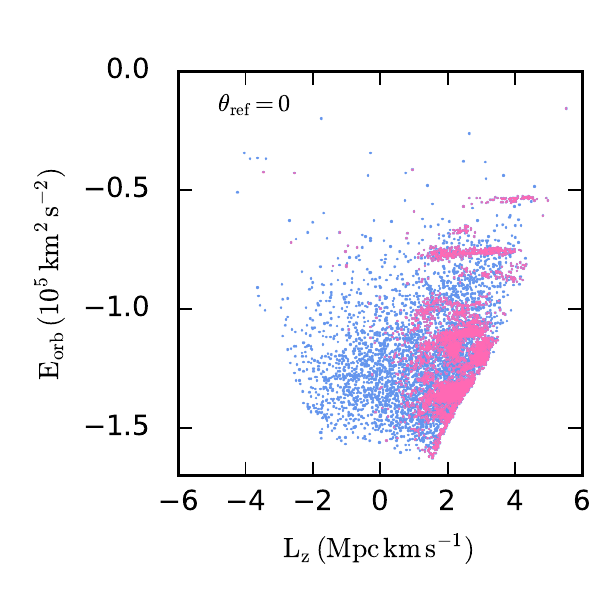}
\includegraphics[scale=0.8,trim={1.6cm 0 0 0.5cm},clip]{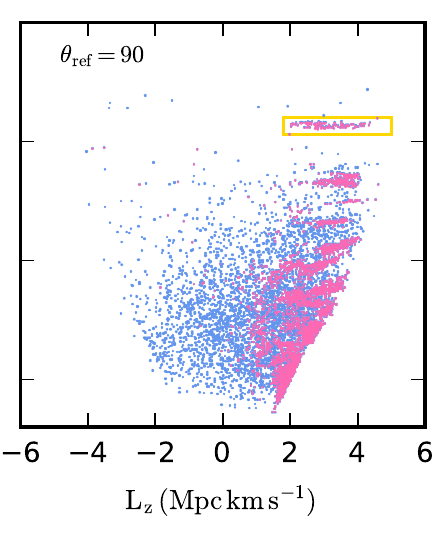}
\includegraphics[scale=0.8,trim={1.6cm 0 0 0.5cm},clip]{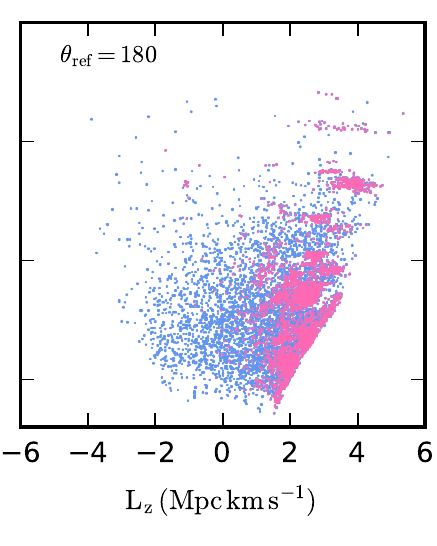}
\includegraphics[scale=0.8,trim={1.6cm 0 0 0.5cm},clip]{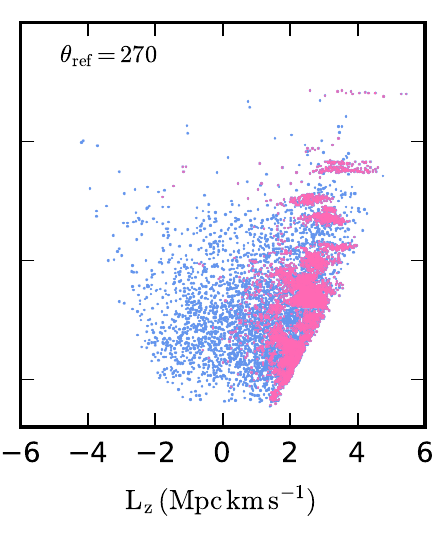}\\
    \caption{{\it Upper panels:} Total velocity distributions for the accreted star particles  selected within a sphere of radius 3 kpc (blue), centred on four different solar positions (from left ot right), in simulation Au 23. The power law distribution for the most likely parameters for the slope, $k$, and escape velocity, $v_{\rm esc}$, is denoted by the solid curve. The true escape velocity is indicated in brackets and by the vertical green line. The empty green histogram shows the velocity distribution for all dark matter particles within each volume. {\it Lower panels:} The distribution of selected accreted star particles in total $E_{\rm orb}$-$L_z$ space. The escape velocity and power law slope (upper panels) vary considerably between different solar positions. This is associated with the clearly visible variation of substructure at different positions (lower panels). In pink, we highlight the material that belonged to the most massive merger. It is clear that most of the phase space substructure derives from this material, as do the many bumps in the velocity distribution.}
    \label{au23}
\end{figure*}

To constrain the dark matter mass, $M_{\rm DM, 200}$, we first calculate the escape speed at the solar radius: $v_e(r_0) = \sqrt{2(\Phi (r_0)-\Phi (2R_{200}))}$, where $\Phi$ is the total gravitational potential from all baryonic and dark material. We assume an NFW profile \citep{NFW97}, and calculate its potential at $r_0$ and $2R_{200}$ for a range of values for $M_{\rm DM,200}$ and  concentration, $c$, that span uniform grids of $\log _{10} M_{\rm DM, 200}\in [11.5,12.5]$ and $c\in [1,30]$. The baryonic contribution to the potential is calculated from the simulation data. We then evaluate the probability distribution of the dark halo parameters using the escape speed derived from the high velocity tail, after marginalising over $k$ and $\gamma$, the latter of which is poorly constrained given the narrow radial range of stars under consideration. To yield a final mass estimate, we combine the probability distribution based on the escape speed with constraints on $M_{\rm DM,200}$ and $c$ from the circular velocity, $v_c (r_0) = \sqrt{GM(<r_0)/r_0}$. 

Our method differs from that of D19 in two main ways: firstly, they model the baryonic component with fixed analytic expressions, whereas we calculate the circular velocity and escape speed from the simulated particle data. Secondly, D19 consider counter-rotating stars from the {\it Gaia} DR2, whereas we consider local, {\it accreted} star particles at four solar positions - each located at a radius of 8 kpc in the disc midplane, and equidistant in galacto-centric azimuth - per simulation. Our selection of accreted star particles is appropriate given that counter-rotating stars are mainly {\it ex-situ}, with the additional benefit of maximising the number of solar vicinities that contain at least 240 accreted star particles within a 3 kpc sphere (or 5 kpc, if a 3 kpc sphere contains fewer particles), which was the number of stars selected by D19 in the same volume.

\begin{figure}
\centering
\includegraphics[scale=0.75,trim={1.1cm 0.cm 0.5cm 0.5cm},clip]{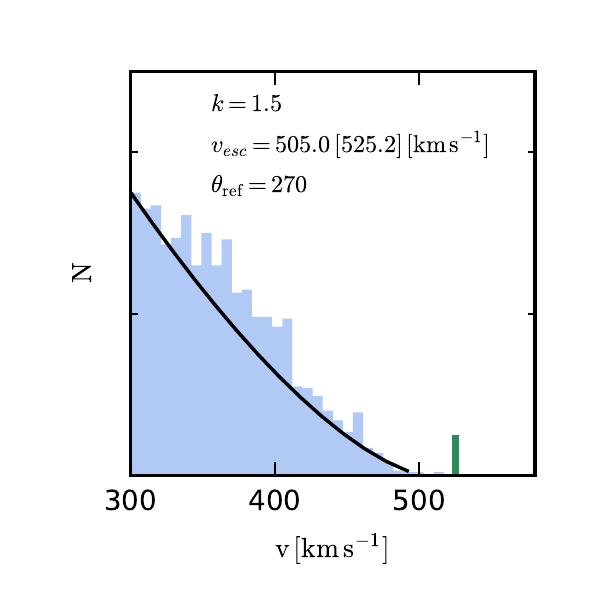}
\includegraphics[scale=0.75,trim={0cm 0cm 0 0.5cm},clip]{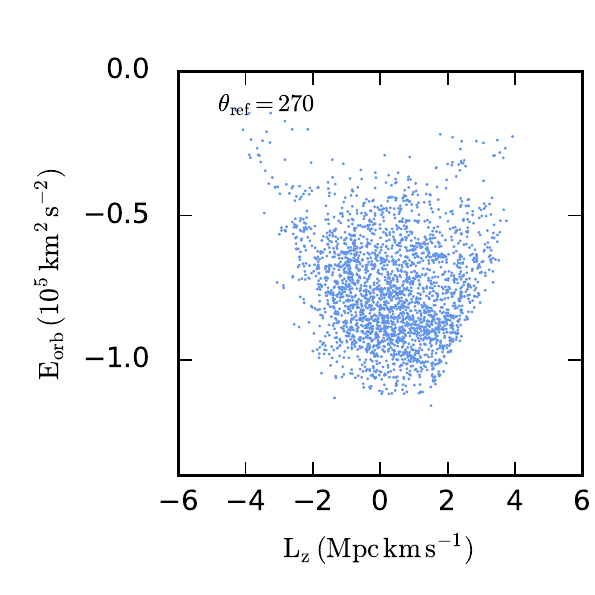}\\
    \caption{As Fig.~\ref{au23}, but for a solar position in simulation Au 6. The derived parameters vary little across all four solar positions, which exhibit relatively smooth phase space distributions. }
    \label{au6}
\end{figure}

\begin{figure*}
\centering
\includegraphics[scale=1.4,trim={0 1.4cm 0 1.1cm},clip]{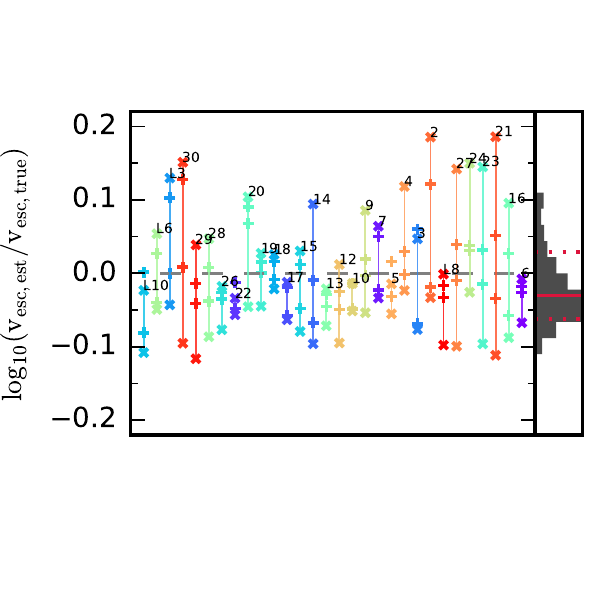}\hspace{0.5cm}
\includegraphics[scale=1.4,trim={0 1.4cm 0 1.1cm},clip]{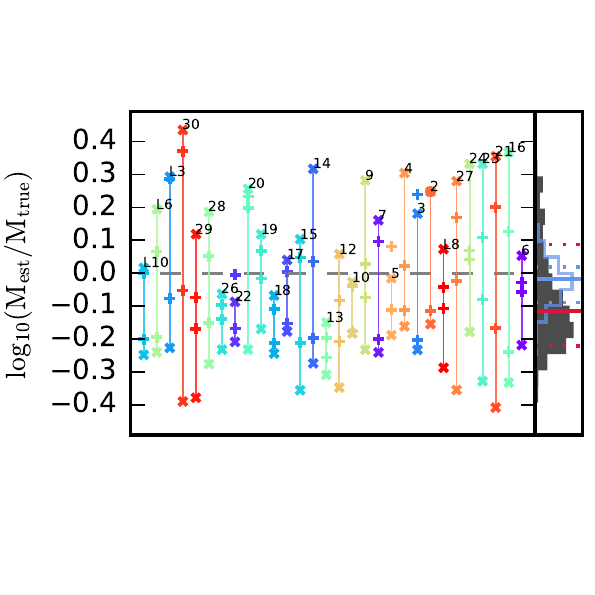}
    \caption{The maximum and minimum values of the logarithm of the estimated values divided by the true values for each simulation. The crosses indicate the maximum and minimum values from the samples of 240 accreted stars, whereas the plusses show the same for each of the fully sampled solar positions per simulation (halo numbers are indicated for each set of positions). The dashed grey line indicates the 1:1 relation. Both escape speed and total mass estimates exhibit significant scatter across the sample, but also within individual haloes. This scatter is larger for the samples of 240 stars (crosses) compared to the full star particle data (plusses). The distributions, median values (solid red lines) and 16th and 84th percentiles (dotted red lines) for the samples of 240 stars are indicated on the right-hand side of each panel: the median estimated escape speed and total halo mass fall below the true values by $\sim 7\%$ and $\sim 20 \%$, and have a spread of $\sim 0.1$ dex and $\sim 0.3$ dex, respectively. The blue empty histogram shows the estimated mass distribution if the true escape speed is used to constrain $M_{\rm DM, 200}$, which results in negligible bias and $\sim 0.1$ dex of scatter.}
    \label{mest}
\end{figure*}

\section{Results}

\subsection{Illustrative cases}

We begin by calculating the true escape speed in each simulation, for which we have the total gravitational potential of each particle available. We calculate the true escape speed from the gravitational potential at $r_0$ and the mean graviational potential of particles in a thin spherical shell at $2R_{\rm 200}$. To assess the accuracy with which the method described above is able to recover the true escape speed from a smooth, well-mixed distribution, we apply the method to all dark matter particles contained within each volume, which are expected to be virialized. The upper panels of Fig.~\ref{au23} show the velocity distributions of these dark matter particles within each volume for simulation Au 23. As expected, the distributions are smooth and are populated all the way up to the true escape speed, which is recovered within 1\%. This verifies that the method works well for smooth distributions that reach the escape speed.

Contrary to the dark matter particles, the upper panels of Fig.~\ref{au23} show that velocity distributions of accreted star particles exhibit multiple bumpy features at both low ($\sim 300$ $\rm km \, s^{-1}$) and high ($> 400$ $\rm km \, s^{-1}$) velocities. The shapes of the distributions, as well as the estimated escape speeds, vary with solar position: the escape speed shows a variation of $\sim 70$ $\rm km \, s^{-1}$, with the lowest and highest estimates bracketing the true escape speed. This variation in escape speed estimates between the different solar positions can be understood from inspection of the distribution of these particles in phase space. We define the total specific energy of a particle as $E_{\rm orb} = 0.5v^2 + \Phi$, where $v$ is the velocity of the particle, $\Phi$ is the gravitational potential. We define the magnitude of the vertical component of the specific angular momentum of particles, in a coordinate frame whose $z$-axis is aligned with the spin-axis of the disc, as $L_z = (\mathbf{x} \times \mathbf{v} ) \cdot \mathbf{\hat{z}}$, where $\mathbf{x}$ and $\mathbf{v}$ are the three dimensional position and velocity of the particle. The lower panels of Fig.~\ref{au23} show the distribution of selected accreted star particles in the $E_{\rm orb}$-$L_z$ plane at each of the solar positions. These diagrams reveal a highly structured phase space distribution, characterised by a variety of distinct stream-like features such as clumps at high binding energies, and narrow structures at low binding energies. The overestimate in the $\theta_{\rm ref}=90$ volume is largely driven by the under-populated velocities in the range $\sim 320$-$400$ $\rm km\,s^{-1}$ relative to the peaks at lower and higher velocities, which skews the likelihood function to estimate too high an escape speed. We find that removing the most weakly bound stream stars in this volume (yellow box in Fig.~\ref{au23}) alleviates the skew, but also removes the highest velocity stars from the distribution, resulting in an  {\it underestimate} of the escape speed. This illustrates the extreme sensitivity of the escape speed estimate to the details of dynamical substructure. 

We emphasise that almost all of the substructure in each of the lower panels of Fig.~\ref{au23} originates from the same progenitor system that merged with the primary galaxy approximately 3 Gyr ago. The peak stellar mass of this object (i.e. the highest stellar mass it attained over its history) is $\sim 5.4 \times10^9$ $\rm M_{\odot}$, and it experienced two pericentric passages before it merged. The abundance and diversity of substructure in this example can therefore be explained by repeated stellar stripping events of this satellite that occur during infall; stars stripped at different times and places, as well as multiple wraps of orbits from a single stripping event, have different binding energies and populate different regions of phase space as streams. The upper panels of  Fig.~\ref{au23} show that this debris is responsible for the bumps in the velocity distribution at each solar position. 

In contrast to highly structured phase space distributions, we show in Fig.~\ref{au6} an example of a smooth local halo star distribution. All four solar volumes appear similarly smooth, and as a result, the high velocity tail is relatively smooth. As such,  the estimated escape speed is consistent across all volumes, although it falls $\sim 20$ $\rm km\,s^{-1}$ short of the true value. This halo has experienced a quiet merger history since $z\sim 1$ relative to the halo shown in Fig.~\ref{au23}.

\subsection{Simulation sample}

We now investigate the estimated escape speeds and total halo masses with respect to their true values for a suite of simulations. We include haloes that contain at least 240 accreted star particles in solar-like vicinities (Section.~\ref{meth}), which results in a total of 30 haloes: 23 at level 4 resolution; and 7 at level 3 resolution. We have verified that the results do not change if the level 3 simulations are replaced with their level 4 counterparts. The number of accreted star particles in each volume varies between several hundred to several thousand across the suite. To match the {\it Gaia} DR2 sample size of D19, we draw random samples of 240 accreted star particles from each full sample volume until they are used up (e.g., a volume with 1000 star particles is sampled 4 times). This approach increases the effective size of our sample of local accreted star volumes to a total of 892, which yields better statistics and incorporates the same degree of noise as in the D19 sample.

In Fig.~\ref{mest}, we show the range of estimated escape speeds (left panel) and total masses (right panel) for the random samples (crosses) for each simulation. The distributions of the escape speed and total mass estimates are similarly smooth and non-Gaussian, with median values of $-0.03$ dex and $-0.11$ dex, respectively. The 16th-84th percentile range is $\sim 0.1$ dex and $\sim 0.3$ dex for the escape speed and total mass, respectively, indicating the sensitivity of the mass estimate to that of the escape speed. The scatter between positions in each halo is comparable to that between simulations, which indicates that noise (from 240 star particles) and variation of substructure between positions is at least as important as the merger history for the estimate accuracy. For many haloes, the range of estimates of both quantities from all accreted particles at each solar position (plusses) is smaller (but significant) than that of the random samples, which illustrates the joint effect of noise and substructure on the scatter. Fig.~\ref{mest} shows also the estimated mass distribution if the true escape speed is used to constrain $M_{\rm DM, 200}$. This shows that the assumption of a spherical NFW profile leads to a negligible bias and a scatter of $\sim 0.1$ dex.

For the full accreted particle data at each position, we discern a range of accuracies for both escape speed and mass that reflect the significant variation in dynamical structure (from smooth to highly structured) both between positions within the same halo and between different haloes. For example, haloes with the smoothest structure (as in Fig.~\ref{au6}) at all positions (e.g., haoles 6, 9 and 26) typically either return good estimates or underestimates with a low degree of scatter between positions. Evidently, the high velocity tail does not extend to the escape speed for the latter. In contrast, haloes with the largest variations (e.g., haloes 16, 21) tend to have prominent dynamical substructure (as in Fig.~\ref{au23}) created by systems of peak stellar mass greater than $10^9$ $\rm M_{\odot}$ that merged within the last several gigayears of evolution after several pericentric passages.


\section{Discussion}

The local stellar halo in the MW is thought to be dominated by a single component, characterised by a radially anisotropic velocity distribution (``the Sausage''), sculpted long ago from a single progenitor \citep[][]{BEE18,HBK18}. \citet{FBD19} found that the \textlcsc{Auriga} haloes with analogous properties formed most of their stellar halo at early times and experienced their last major merger between 6 and 10 Gyr ago (haloes 5, 9, 10, 17, 18, 22, 26). For our full local halo samples (crosses in Fig.~\ref{mest}), the mass estimates are consistently within $\sim 10\%$ of the true mass at each position for some haloes (6, 9), whereas others produce mass estimates up to $\sim40\%$ different from (usually lower than) the true mass. The latter are consistent with the high velocity tail truncating below the true escape speed. This may be understood from the orbits of early merger ($z>1$) progenitors, which need to be sufficiently tight to be captured by the main galaxy, such that its stars are not expected to go out as far as today's $2R_{200}$ after disruption. This is consistent with the mass estimates in this work falling $20\%$ below, on average, the true mass. This reasoning applies also to a similar underestimate reported in Sec. 5.2 of \citet{WE99} related to streams that fall short of the escape speed.

With respect to dynamical substructure, the ``Sausage'' haloes with the largest spreads in Fig.~\ref{mest} (5, 10) do contain moderate substructure at some positions, despite their early halo assembly. More importantly, the continual discovery of dynamical substructure of halo stars in the local vicinity \citep[e.g.][]{HWZ99,HVB17,MEB18a,MEB18b,KHV18,IMM19} indicates that the scatter (and therefore, uncertainty) in the mass measurements linked to streams found in this paper are an important consideration for the MW. This is pertinent for future larger volume samples, especially if they include any member stars of the richly structured Sagittarius stream \citep[][]{SHD17}. It is therefore clear that, in addition to larger and more accurate future datasets, a complete mapping of the streams and assembly history of the stellar halo is required to fully understand the extent of these biases. For the present situation, we may use the biases found in Fig.~\ref{mest} (median and spread) to revise the MW mass derived in D19 to $1.29 ^{+0.37}_{-0.47} \times 10^{12}$ $\rm M_{\odot}$, which agrees well with estimates based on satellites \citep[][]{CCD19} and globular clusters \citep[][]{PH19}.

\section{Conclusions}

We have used a total of 30 magneto-hydrodynamical cosmological zoom-in simulations for the formation of MW-mass galaxies to investigate the connection between the phase-space substructure and high velocity tail of accreted stars in the vicinity of solar-like positions. We investigate how this substructure and velocity distribution varies between different solar-like positions and across the simulation set. We have applied a common method for estimating the escape speed and total mass of each halo from the high velocity tail D19. Our conclusions are summarised as follows:

\begin{itemize}
    \item The properties of dynamical substructure depend on the merger history and vary dramatically (from smooth to highly structured) between positions in individual haloes and between different haloes. The most significant phase space substructure originates from massive progenitors, that had: peak stellar masses $\gtrsim 10^9$ $\rm M_{\odot}$; and several pericentric passages before merging at late times, whereas the quietest merger histories have the smoothest phase space structure. 
    
    \item The method to estimate the escape velocity and halo mass from the high-velocity tail is extremely sensitive to the dynamical structure of halo stars.  Dynamical substructure (streams) leads to un-evenly populated velocity distributions that act to increase the scatter of mass estimates between positions within individual haloes. Smooth phase space distributions are conducive to better estimates, although they have a tendency to underestimate the mass owing to the truncation of the high velocity tail below the true escape speed. However, the effects of noise and streams lead to significant scatter in the estimated quantities, even in cases for which the velocity distribution is relatively smooth.
                
    \item The distribution of escape speed and mass estimates from all simulations and positions have medians lower than the true values by $7\%$ and $20 \%$, and spreads of factors of $\sim1.3$ and $\sim 2$, respectively. Based on these biases, the MW mass estimate derived in D19 is revised to $1.29 ^{+0.37}_{-0.47} \times 10^{12}$ $\rm M_{\odot}$.
\end{itemize}

\section*{Acknowledgements}
RG thanks Guinevere Kauffmann, Wilma Trick and Vasily Belokurov for useful comments. AD is supported by a Royal Society University Research Fellow-ship. AD also acknowledges the support from the STFC grant ST/P000541/1. FAG acknowledges financial support from CONICYT through the project FONDECYT Regular Nr. 1181264, and funding from the Max Planck Society through a Partner Group grant. FM is supported through the Program ``Rita Levi Montalcini'' of the Italian MIUR.

\bibliographystyle{mnras}
\bibliography{mnras_template.bbl}

\bsp	
\label{lastpage}
\end{document}